\documentstyle[11pt,aaspp]{article}
\begin{document}

\newbox\grsign \setbox\grsign=\hbox{$>$} \newdimen\grdimen \grdimen=\ht\grsign
\newbox\simlessbox \newbox\simgreatbox
\setbox\simgreatbox=\hbox{\raise.5ex\hbox{$>$}\llap
     {\lower.5ex\hbox{$\sim$}}}\ht1=\grdimen\dp1=0pt
\setbox\simlessbox=\hbox{\raise.5ex\hbox{$<$}\llap
     {\lower.5ex\hbox{$\sim$}}}\ht2=\grdimen\dp2=0pt
\def\gtorder{\mathrel{\copy\simgreatbox}}
\def\ltorder{\mathrel{\copy\simlessbox}}
\def\simgreat{\mathrel{\copy\simgreatbox}}
\def\simless{\mathrel{\copy\simlessbox}}

\def\chaphead{}

\def\hut{Hubble type\ }
\def\vc{V$_{\rm C}$\ }
\def\mb{M$_{\rm B}$\ }
\def\av{A$_{\rm V}$\ }
\def\lamlam{$\lambda\lambda$}

\def\deg{$^\circ$}
\def\degrees{$^\circ$}
\def\Vlasov{collisionless Boltzmann\ }
\def\lsls{\ll}
\def\grgr{\gg}
\def\erf{\mathop{\rm erf}\nolimits} 
\def\eqv{\equiv}
\def\real{\Re e}
\def\imag{\Im m}
\def\ctrline#1{\centerline{#1}}
\def\spose#1{\hbox to 0pt{#1\hss}}
     
\def\={\overline}
\def\sections{\S}
\newcount\notenumber
\notenumber=1
\newcount\eqnumber
\eqnumber=1
\newcount\fignumber
\fignumber=1
\newbox\abstr
\newbox\figca     
\def\yyskip{\penalty-100\vskip6pt plus6pt minus4pt}
     
\def\numberpara{\yyskip\noindent}
     
\def\km{{\rm\,km}}
\def\kms{{\rm\ km\ s$^{-1}$}}
\def\kpc{{\rm\,kpc}}
\def\mpc{{\rm\,Mpc}}
\def\etal{{\it et al. }}
\def\eg{{\it e.g. }}
\def\ie{{\it i.e. }}
\def\cf{{\it cf. }}
\def\msun{{\rm\,M_\odot}}
\def\lsun{{\rm\,L_\odot}}
\def\rsun{{\rm\,R_\odot}}
\def\pc{{\rm\,pc}}
\def\cm{{\rm\,cm}}
\def\yr{{\rm\,yr}}
\def\au{{\rm\,AU}}
\def\AU{{\rm\,AU}}
\def\gm{{\rm\,g}}
\def\s{{\rmss}}
\def\dyne{{\rm\,dyne}}
     
\def\note#1{\footnote{$^{\the\notenumber}$}{#1}\global\advance\notenumber by 1}
\def\foot#1{\raise3pt\hbox{\eightrm \the\notenumber}
     \hfil\par\vskip3pt\hrule\vskip6pt
     \noindent\raise3pt\hbox{\eightrm \the\notenumber}
     #1\par\vskip6pt\hrule\vskip3pt\noindent\global\advance\notenumber by 1}
\def\propo{\propto}
\def\larrow{\leftarrow}
\def\rarrow{\rightarrow}
\def\sectionhead#1{\penalty-200\vskip24pt plus12pt minus6pt
        \centerline{\bbrm#1}\vskip6pt}
     
\def\Dt{\spose{\raise 1.5ex\hbox{\hskip3pt$\mathchar"201$}}}    
\def\dt{\spose{\raise 1.0ex\hbox{\hskip2pt$\mathchar"201$}}}    
\def\llangle{\langle\langle}
\def\rrangle{\rangle\rangle}
\def\ldotss{\ldots}
\def\del{\b\nabla}
     
\def\new{{\rm\chaphead\the\eqnumber}\global\advance\eqnumber by 1}
\def\ref#1{\advance\eqnumber by -#1 \chaphead\the\eqnumber
     \advance\eqnumber by #1 }
\def\last{\advance\eqnumber by -1 {\rm\chaphead\the\eqnumber}\advance
     \eqnumber by 1}
\def\eqnam#1{\xdef#1{\chaphead\the\eqnumber}}
     
\def\nfig{\chaphead\the\fignumber\global\advance\fignumber by 1}
\def\nfiga#1{\chaphead\the\fignumber{#1}\global\advance\fignumber by 1}
\def\rfig#1{\advance\fignumber by -#1 \chaphead\the\fignumber
     \advance\fignumber by #1}
\def\refindent{\par\noindent\parskip=3pt\hangindent=3pc\hangafter=1 }

\def\apj#1#2#3{\refindent#1,  {ApJ,\ }{\bf#2}, #3}
\def\apjsup#1#2#3{\refindent#1,  {ApJS,\ }{\bf#2}, #3}
\def\aasup#1#2#3{\refindent#1,  { A \& AS\ }{\bf#2}, #3}
\def\aas#1#2#3{\refindent#1,  { Bull. Am. Astr. Soc.,\ }{\bf#2}, #3}
\def\apjlett#1#2#3{\refindent#1,  { ApJL,\  }{\bf#2}, #3}
\def\mn#1#2#3{\refindent#1,  { MNRAS,\ }{\bf#2}, #3}
\def\mnras#1#2#3{\refindent#1,  { M.N.R.A.S., }{\bf#2}, #3}
\def\annrev#1#2#3{\refindent#1, { ARA \& A,\ }
{\bf2}, #3}
\def\aj#1#2#3{\refindent#1,  { AJ,\  }{\bf#2}, #3}
\def\phrev#1#2#3{\refindent#1, { Phys. Rev.,}{\bf#2}, #3}
\def\aa#1#2#3{\refindent#1,  { A \& A,\ }{\bf#2}, #3}
\def\nature#1#2#3{\refindent#1,  { Nature,\ }{\bf#2}, #3}
\def\icarus#1#2#3{\refindent#1,  { Icarus, }{\bf#2}, #3}
\def\pasp#1#2#3{\refindent#1,  { PASP,\ }{\bf#2}, #3}
\def\appopt#1#2#3{\refindent#1,  { App. Optics,\  }{\bf#2}, #3}
\def\spie#1#2#3{\refindent#1,  { Proc. of SPIE,\  }{\bf#2}, #3}
\def\opteng#1#2#3{\refindent#1,  { Opt. Eng.,\  }{\bf#2}, #3}
\def\refpaper#1#2#3#4{\refindent#1,  { #2 }{\bf#3}, #4}
\def\refbook#1{\refindent#1}
\def\science#1#2#3{\refindent#1, { Science, }{\bf#2}, #3}
     
\def\chapbegin#1#2{\eject\vskip36pt\par\noindent{\chapheadfont#1\hskip30pt
     #2}\vskip36pt}
\def\sectionbegin#1{\vskip30pt\par\noindent{\bf#1}\par\vskip15pt}
\def\subsectionbegin#1{\vskip20pt\par\noindent{\bf#1}\par\vskip12pt}
\def\topic#1{\vskip5pt\par\noindent{\topicfont#1}\ \ \ \ \ }
     
\def\ltsim{\mathrel{\spose{\lower 3pt\hbox{$\mathchar"218$}}
     \raise 2.0pt\hbox{$\mathchar"13C$}}}
\def\gtsim{\mathrel{\spose{\lower 3pt\hbox{$\mathchar"218$}}
     \raise 2.0pt\hbox{$\mathchar"13E$}}}
     
\def\sec{\hbox{$^s$\hskip-3pt .}}
\def\gg{\hbox{$>$\hskip-4pt $>$}}
\parskip=3pt
\def\gapprox{$_ >\atop{^\sim}$}     
\def\lapprox{$_ <\atop{^\sim}$}     
\def\apequal{\mathrel{\spose{\lower 1pt\hbox{$\mathchar"218$}}
     \raise 2.0pt\hbox{$\mathchar"218$}}}

\def\oforder{$\sim$} \def\inv{$^{-1}$}
\def\>={$\geq$} \def\<={$\leq$} \def\ks{km s\inv} \def\kms{km s\inv}
\def\lith{$h$} \def\sig{$\sigma$} \def\sigp{$\sigma^{\prime}_r$}
\def\meanz{$\overline \upsilon$} \def\nc{$N_c$} \def\rc{$r_c$}
\def\twidle{$\sim$} \def\sigmar{$\sigma_r$}
\def\Mstar{{M_B}^*}
\def\Mdot{M_{\odot}}

\title{HIERARCHICAL EVOLUTION IN POOR GROUPS OF GALAXIES}  
\author{Ann I. Zabludoff\altaffilmark{1,2} 
and John S. Mulchaey\altaffilmark{1}}
\altaffiltext{1}{Observatories of the
Carnegie Institution of Washington, 813 Santa Barbara St., Pasadena,
CA 91101, E-mail: mulchaey@pegasus.ociw.edu}
\altaffiltext{2}{UCO/Lick Observatory and Board of Astronomy and
Astrophysics, University of California at Santa Cruz, Santa Cruz, CA,
95064, E-mail: aiz@ucolick.org}

\centerline {Accepted for publication in {\it Astrophysical Journal Letters}}

\singlespace
\abstract{
To examine the evidence for hierarchical evolution on mass scales of
$\sim 10^{13}$-$10^{14} \Mdot$,
we apply a statistic that measures correlations between
galaxy velocity and projected position (Dressler \& Shectman 1988)
to data for six poor groups of galaxies, 
HCG 42, HCG 62, NGC 533, NGC 2563, NGC 5129, and
NGC 741.  Each group has more than 30 identified members (Zabludoff \&
Mulchaey 1998ab).  The statistic is sensitive to clumps of galaxies 
on the sky whose mean velocity and velocity dispersion deviate from
the kinematics of the group as a whole.  The
kinematics of galaxies within $\sim 0.1$\lith\inv\ Mpc of
the group center do not deviate from the global values,
supporting our earlier claim that
the group cores are close to virialization or virialized.  We detect
significant substructure (at $\geq 99.9$\% confidence) in
the two groups with the most confirmed members, HCG 62 and NGC 741, 
that is attributable mostly
to a subgroup lying $\sim 0.3$-0.4\lith\inv\ Mpc outside of the core.
We conclude that at least some poor groups,
like rich clusters, are evolving via the accretion of smaller
structures from the field.  
With larger poor group
surveys, the incidence of such accretion and the distribution of
subgroup masses are potential constraints of cosmological models
on mass scales of $\simless 10^{13}$-$10^{14} \Mdot$ and on physical scales 
of $\simless 0.5$\lith\inv\ Mpc.

\vskip 0.5cm
\noindent{\it Subject headings}:  galaxies: clustering ---
cosmology: large-scale structure of Universe
}

\vfill\eject
\section{Introduction}
The evolution of structure on different mass scales is one of the
outstanding issues in cosmology.  For example, although galaxy
clusters of $\sim 10^{15} \Mdot$ (including Virgo (Binggeli 1993;
Bohringer \etal 1994), Coma (Mellier \etal 1988; Briel \etal 1992;
White \etal 1993), and Abell 754 (Zabludoff \& Zaritsky 1995)) are
clearly evolving from the accretion of smaller groups, it is uncertain
whether poor groups of $\sim 10^{13}$-$10^{14} \Mdot$ also evolve
hierarchically.  There is some indirect evidence that the evolution of
poor groups is similar to rich clusters; the galaxies and hot gas in poor
groups follow the same relationships found among the X-ray
temperature, X-ray luminosity, and galaxy velocity dispersion for rich
clusters (Mulchaey \& Zabludoff 1998, hereafter MZ98).  Historically,
however, the number of known poor group members has been too small
to examine individual groups for direct
evidence of hierarchical evolution.

Multi-object spectroscopy now makes it possible to obtain
``cluster-size'' samples of galaxies in poor groups and to identify
substructure, if it exists, in the same manner as for rich clusters.
Substructure in clusters was not detected until the number of
spectroscopically-confirmed cluster members exceeded $\sim 30$-50
galaxies.  Recent poor group surveys have reached this membership
level (Zabludoff \& Mulchaey 1998ab; hereafter ZM98a and b).  The
discovery of substructure in poor groups would provide new
cosmological constraints by establishing that hierarchical evolution
is occurring on mass scales of $\sim 10^{13}$-$10^{14} \Mdot$ and
on physical scales of $\sim 0.5$\lith\inv\ Mpc.  The detection of
substructure would also support the picture in which at least some
poor groups evolve as low-mass analogs to rich clusters.

In this Letter, we describe the results from applying a substructure
statistic (Dressler \& Shectman 1988; hereafter DS88) 
to the six best sampled poor groups in ZM98ab, the
most detailed spectroscopic survey of poor groups to date.

\section{The Group Sample}
In this analysis of substructure in poor groups, we consider six
nearby systems ($3800 < cz < 7000$ \ks) drawn from our spectroscopic
survey of 12 optically-identified groups (ZM98a).  We choose the six
groups, HCG 42, HCG 62, NGC 533, NGC 741, NGC 2563, and NGC 5129,
because each has more than 30 spectroscopically-confirmed members
(see ZM98a for the details of the membership algorithm).
The group members range in absolute magnitude from the three or four
brightest ($M \simless M^*$) galaxies identified in past surveys to the large
population of dwarfs ($M \simless M^* + 4$) discovered in ZM98a.  All
six groups have X-ray halos extending out to radii of $\sim 200$-300
kpc and X-ray temperatures of $\sim 1$ keV (MZ98).  In this Letter, we
use the galaxy velocity data from ZM98a combined with new velocities
for 51 additional group members (ZM98b).  The number of known members
within $\sim 0.5$\lith\inv\ Mpc ($\sim$ the virial radius) 
of each group center ranges from 35 to 63 galaxies.

\section{Results and Discussion}
To search for substructure in the six poor groups, we use the method
applied by DS88 to rich clusters.  This test identifies a fixed number
of nearest neighbors on the sky around each galaxy, 
calculates the local mean
velocity and velocity dispersion of the subsample, and compares these
values with the mean velocity and velocity dispersion of the entire
group.  The kinematic deviations of the subsamples
from the global values are summed.  This sum is larger for a group with
a kinematically distinct subgroup than for a similar group without
substructure.  

For each galaxy $i$, the deviation of its nearest projected neighbors
from the kinematics of the group as a whole is defined as
${\delta_i} \equiv (n^{1/2}/ \sigma_r) \lbrack 
(\upsilon_{loc} - \overline \upsilon)^2
+ (\sigma_{r,loc} - \sigma_r)^2 \rbrack ^{1/2}$,
where $\overline \upsilon$ is the mean velocity for the group,
$\sigma_r$ is the group velocity dispersion, 
and $n$ is the number of nearest neighbors (including the galaxy)
used to determine the local mean velocity
$\overline \upsilon_{loc}$ and local velocity dispersion $\sigma_{r,loc}$.
The total deviation for the group is defined as the sum of the local
deviations,
$\Delta \equiv \sum |\delta_i|$ for all $i \leq N_{grp}$, the number of 
group members.
As pointed out in DS88, the
$\Delta$ statistic is similar to the $\chi^2$ statistic, except that the
$\delta_i$'s are not squared before summation in order to reduce
the contributions of the largest, rarest deviations.
If the galaxy velocity distribution of the group is close
to Gaussian, and the local variations are only random fluctuations, 
$\Delta \simeq N_{grp}$.  

To calculate $\delta$, we choose $n = 11$ (as in DS88).
This choice allows robust determinations of $\overline
\upsilon_{r,loc}$ and $\sigma_{r,loc}$.  Silverman (1986) argues that
using $n \sim {N_{grp}}^{1/2}$ nearest neighbors (= 6-8 for these groups)
maximizes the sensitivity of such a test to small scale structures
while reducing its sensitivity to fluctuations within the Poisson
noise (also see Bird 1994b).  
To check the robustness of the $n=11$ assumption, we compare
the results below with those for $n=6$.  The conclusions drawn from the
$n=6$ and $n=11$ cases are the same.

Calibration of the $\Delta$ statistic for each group
is required because 1) the $\delta_i$'s
are not statstically independent and 2) the velocity distribution may
not be intrisically Gaussian even if there are no subgroups
({\it e.g.}, the group members may follow
predominantly circular or radial orbits).
We determine the significance of the observed $\Delta$ by comparing it
with the results of 1000 Monte Carlo trials in which
galaxy velocities are drawn randomly from
the observed distribution and assigned to galaxy positions.  This
scrambling technique effectively destroys any substructure (DS88).
If the probability is low that a group without substructure has a
$\Delta$ value at least as 
large as that observed, then we consider the substructure detection
significant.

In two groups, HCG 62 and NGC 741, the observed value of $\Delta$ is
significant at the $\geq 99.9\%$ confidence level.  
Such high $\Delta$ values might arise from substructure, but
also could result 
from smooth variations in the group's velocity field ({\it e.g.},
rotation or velocity shear (Malumuth \etal 1992))
and/or from a dependence of $\sigma_r$ on radius (Bird 1994b).
ZM98a show that $\sigma_r$ is constant
out to radii of $\sim 0.5$\lith\inv\ Mpc in the combined
velocity dispersion profile for the sample groups.
To determine whether substructure is in fact responsible for the high
$\Delta$ values in HCG 62 and NGC 741, we examine 
the local deviation $\delta$ for each group member.
A concentration of large $\delta$ values on the sky indicates
a kinematically distinct subgroup.

Figure 1 shows the projected spatial
distribution of group members for each group (top panel).
The second panel shows this distribution with
the radii of the circles 
weighted by $e^\delta$ (as in DS88).
Because each point is not statistically independent, a few very deviant 
galaxies can boost the $\delta$ values for a
large number of nearby points.  Therefore, a visual comparison with the
Monte Carlo simulations is required to assess the significance of
any structures.
The third panel shows the results of the Monte Carlo trial (out of 1000) 
with the largest $\Delta$ value, or greatest total deviation. 
The results of the trial with the median $\Delta$
value are in the bottom panel.  

The significance of the
seven large clustered circles
to the northeast of HCG 62 and five large clustered circles to the south
of NGC 741 is high.  In each case, the large $\delta$ values show
that a subgroup not in equilibrium with the global group potential is the 
principal source of the significant $\Delta$ value.
Each of the two subgroups lies a projected distance of
$\sim 0.3$-0.4\lith\inv\ Mpc outside of the group core.

On the other hand, the clustering of
small circles within $\sim 0.1$\lith\inv\ Mpc of all of the group centers 
indicates that the core mean velocity and velocity dispersion
are similar to the global values for the group.  This result suggests
that the group cores are close to virialization or virialized
and is consistent with the conclusions
from our earlier studies of group dynamics (ZM98a, MZ98).

\section{Conclusions}
Of the six best sampled poor groups of galaxies (each has more than
30 known members), two have significant substructure.
In each case, the substructure detection is due mostly to a subgroup
lying $\sim 0.3$-0.4\lith\inv\ Mpc outside of the core.  It is worth noting
that we detect substructure in the two groups, HCG 62 and NGC 741,
with the most confirmed members and also that our
substructure statistic (from DS88)
is not sensitive to subgroups that are well-superposed on the sky 
(Bird 1993, 1994a).
These two points suggest that the
fraction of poor groups with substructure is higher than
the $\sim 30$\% reported here.
We conclude that poor groups, like rich clusters, evolve hierarchically
and that some poor groups are still accreting smaller structures today
\footnote{Even the Local Group, which is less massive
than these groups ($\sim 4 \times 10^{12} \Mdot$ [Zaritsky \etal 1989] 
versus $\sim 10^{13}$-$10^{14} \Mdot$ [ZM98a]), 
is forming from the collapse of the Milky Way
and M 31 subgroups (Kahn \& Woltjer 1959).  There is also some evidence
that AWM/MKW poor clusters, which are typically $\sim 2$-4 times
more massive than these groups, have bound clumps at much larger radii
($\simgreat 1$-2\lith\inv\ Mpc; Beers \etal 1995).}.
In contrast, the cores of all of the poor groups (within
$\sim 0.1$\lith\inv\ Mpc) appear to be relaxed, an observation
consistent with our earlier claims that the cores
are close to virialization or virialized (ZM98a, MZ98).
Deep spectroscopic surveys of groups in a
{\it large} group sample would place limits on both
the incidence of substructure and the distribution of 
subgroup masses, providing a new test of cosmological models
on mass scales of $\simless 10^{13}$-$10^{14} \Mdot$ and on physical scales 
of $\simless 0.5$\lith\inv\ Mpc.

\vskip 0.3in\noindent 
We thank Dennis Zaritsky
for his comments on the text.
AIZ acknowledges support from the Carnegie and Dudley
Observatories, the AAS, NSF grant AST-95-29259, and NASA grant
HF-01087.01-96A.
JM acknowledges support provided by NASA grant NAG 5-2831 and NAG 5-3529.

\vfill\eject
\centerline{\bf References}

\aj{Beers, T.C., Kriessler, J.R., Bird, C.M., \& Huchra, 
J.P. 1995}{109}{874}
\aa{Binggeli, B. 1993}{98}{275}
\refbook{Bird, C. 1993, Ph.D. Thesis, University of Minnesota and Michigan
State University.}
\aj{Bird, C. 1994a}{107}{1637}
\aj{Bird, C. 1994b}{422}{480}
\nature{Bohringer, H., Briel, U.G., Schwarz, R.A., Voges, W., 
Hartner, G. \& Trumper, J. 1994}{368}{828}
\aa{Briel, U.G., Henry, J.P., \& Bohringer, H. 1992}{259}{L31}
\aj{Dressler, A. \& Shectman, S. 1988}{95}{985}  (DS88)
\apj{Kahn, F.D. \& Woltjer, L. 1959}{130}{705}
\aj{Malumuth, E.M., Kriss, G.A., Dix, W.V.D., Ferguson, H.C., \& Ritchie, C.
1992}{104}{495}
\aa{Mellier, Y., Mathez, G., Mazure, A., Chauvineau, B., Proust, D. 1998}
{199}{67}
\refbook{Mulchaey, J. S. \& Zabludoff, A.I. 1998, ApJ, in press. (MZ98)}
\refbook{Silverman, B. 1986, Density Estimation for Statistics and Data 
Analysis (London: Chapman and Hall)}
\mnras{White, S.D.M., Briel, U., \& Henry, J.P. 1993}{261}{L8}
\apjlett{Zabludoff, A.I. \& Zaritsky, D. 1995}{447}{L21}
\refbook{Zabludoff, A.I. \& Mulchaey, J.S. 1998a, ApJ, in press. (ZM98a)}
\refbook{Zabludoff, A.I. \& Mulchaey, J.S. 1998b, in prep. (ZM98b)}
\apj{Zaritsky, D., Olszewski, E.W., Schommer, R.A., Peterson, R.C., \&
Aaronson, M. 1989}{345}{759}
\vfill\eject

\centerline{\bf Figure Captions}

\noindent
{\bf Figure 1:}
For each group, the top panel shows the distribution of group members
on the sky.  There are 35 galaxies in HCG 42, 63 in HCG 62, 36 in NGC
533, 41 in NGC 741, 44 in NGC 2563, and 38 in NGC 5129.  The dashed
line in the NGC 741 and NGC 5129 fields is the survey boundary (see
ZM98a).  The second panel is the same as the top
panel, except that the radii of the circles are scaled by $e^\delta$,
where $\delta$ is the deviation of the nearest neighbors
from the kinematics of the group as a whole.
The third panel shows the results of the Monte Carlo
trial (out of 1000) with the largest value of the 
Dressler-Shectman (DS88) statistic $\Delta$.  The bottom panel is the trial
with the median value of $\Delta$.  The scale bar below the bottom
panel is 0.5\lith\inv\ Mpc.
\eject
\begin{figure}
\plotone{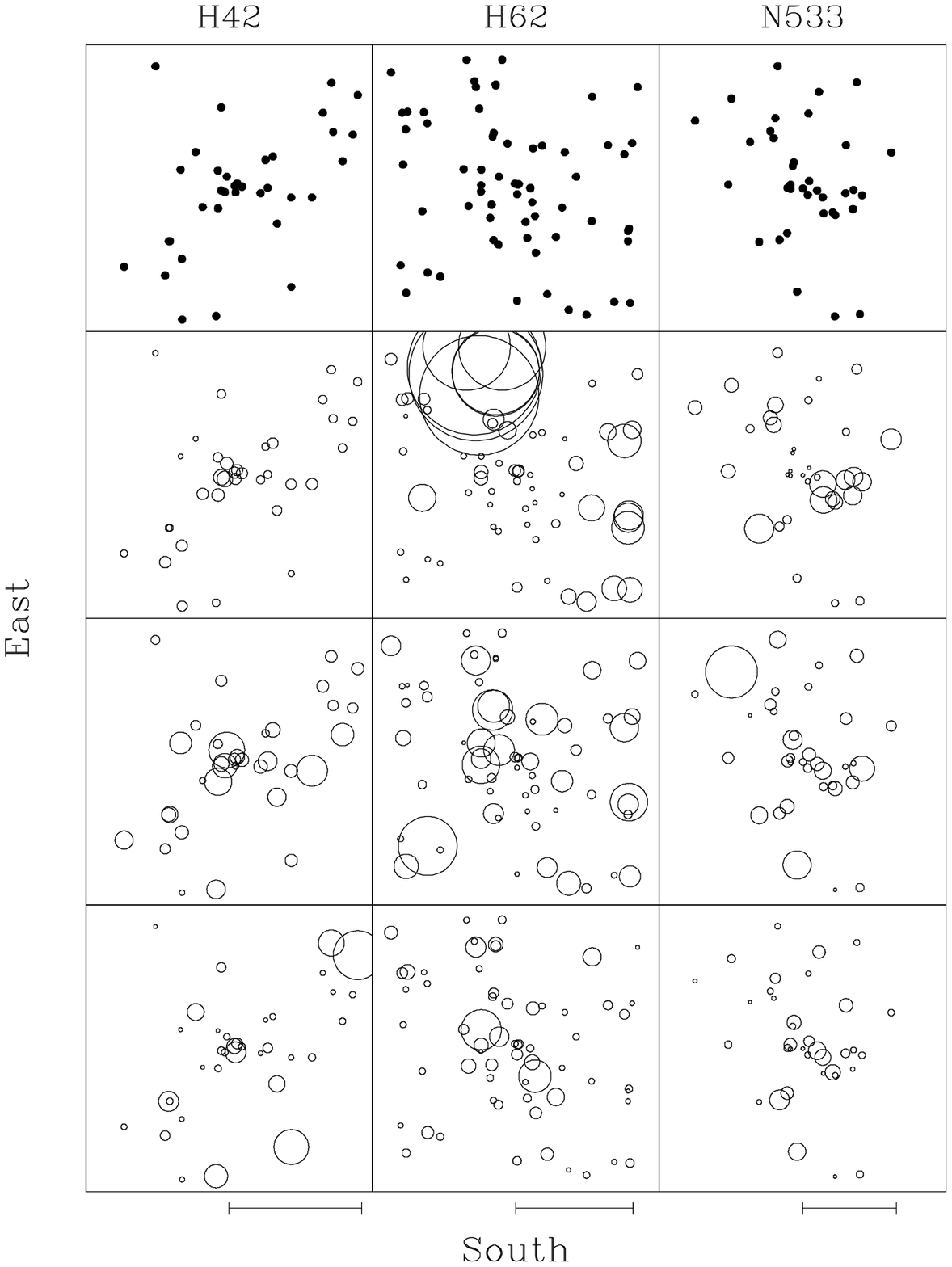}
\caption{}
\end{figure}
\clearpage
\vfill\eject
\clearpage
\setcounter{figure}{0}
\begin{figure}
\plotone{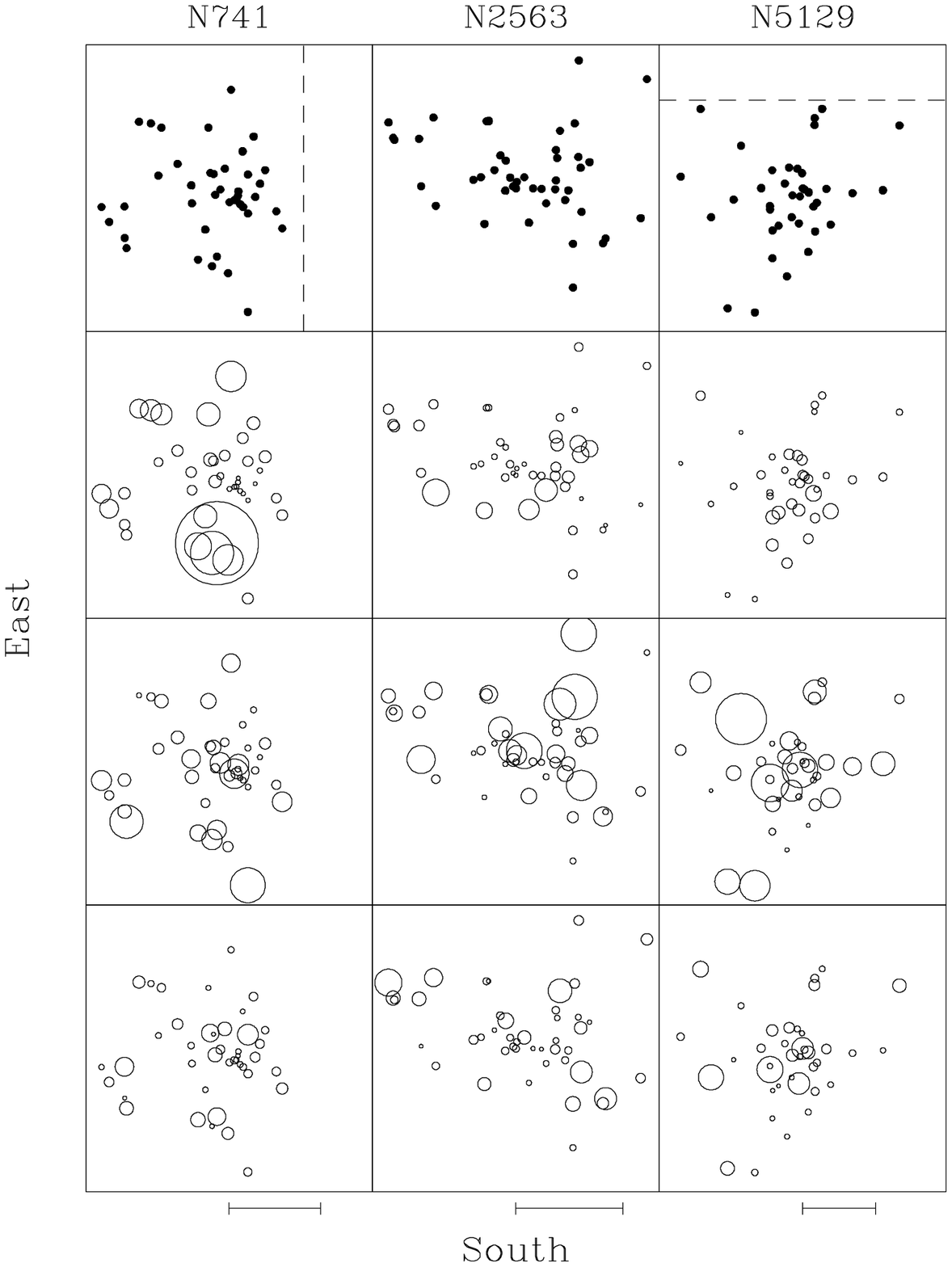}
\caption{{\it cont. }}
\end{figure}
\clearpage
\vfill\eject
\end{document}